\documentclass[12pt]{article}
\usepackage{times}
\usepackage{geometry}
\usepackage[utf8]{inputenc}
\usepackage{enumitem,amssymb}
\usepackage{ragged2e}
\newlist{thematic}{itemize}{8}
\setlist[thematic]{label=$\square$}
\usepackage{pifont}

\usepackage{wrapfig}
\usepackage{graphicx}
\usepackage{caption}
\usepackage{natbib}
\usepackage[usenames, dvipsnames]{xcolor}

\newcommand{\micron}{$\mu\textrm{m}$}

\newcommand{\apj}{ApJ}
\newcommand{\aj}{AJ}
\newcommand{\apjs}{ApJS}

\newcommand{\aap}{A\&A}
\newcommand{\mnras}{MNRAS}

\newcommand{\kemp}[1]{\textcolor{OliveGreen}{#1}}
\newcommand{\kempviolet}[1]{\textcolor{Violet}{#1}}
\newcommand{\kempred}[1]{\textcolor{Red}{#1}}

\begin{document}
\raggedright
\huge
Astro2020 Science White Paper \linebreak

Interstellar Dust Grains: \\
Ultraviolet and Mid-IR Extinction Curves \linebreak
\normalsize

\noindent \textbf{Thematic Areas:} \hspace*{60pt} $\square$ Planetary Systems \hspace*{10pt} $\boxtimes$ Star and Planet Formation \hspace*{20pt}\linebreak
$\square$ Formation and Evolution of Compact Objects \hspace*{31pt} $\square$ Cosmology and Fundamental Physics \linebreak
  $\square$  Stars and Stellar Evolution \hspace*{1pt} $\boxtimes$ Resolved Stellar Populations and their Environments \hspace*{40pt} \linebreak
  $\square$    Galaxy Evolution   \hspace*{45pt} $\square$             Multi-Messenger Astronomy and Astrophysics \hspace*{65pt} \linebreak
  
\textbf{Principal Author:}

Name:	Karl D.\ Gordon
 \linebreak						
Institution:  Space Telescope Science Institute
 \linebreak
Email: kgordon@stsci.edu
 \linebreak
Phone: 410-338-5031
 \linebreak
 
\textbf{Co-authors:} \\
Karl Misselt (Univ.\ of AZ), Yvonne Pendleton (NASA/Ames), Benne Holwerda (Univ.\ of Louisville), Christopher Clark (STScI ), Geoffrey Clayton (Louisiana State Univ.), Lea Hagen (STScI), Julia Roman-Duval (STScI), Adolf Witt (Univ.\ of Toledo), Michael Wolff (Space Science Institute) \linebreak


\textbf{Abstract:}

Interstellar dust plays a central role in shaping the detailed structure of the interstellar medium, thus strongly influencing star formation and galaxy evolution.
Dust extinction provides one of the main pillars of our understanding of interstellar dust while also often being one of the limiting factors when interpreting observations of distant objects, including resolved and unresolved galaxies.
The ultraviolet (UV) and mid-infrared (MIR) wavelength regimes exhibit features of the main components of dust, carbonaceous and silicate materials, and therefore provide the most fruitful avenue for detailed extinction curve studies.
Our current picture of extinction curves is strongly biased to nearby regions in the Milky Way.
The small number of UV extinction curves measured in the Local Group (mainly Magellanic Clouds) clearly indicates that the range of dust properties is significantly broader than those inferred from the UV extinction characteristics of local regions of the  Milky Way.
Obtaining statistically significant samples of UV and MIR extinction measurements for all the dusty Local Group galaxies will provide, for the first time, a basis for understanding dust grains over a wide range of environments.
Obtaining such observations requires sensitive medium-band UV, blue-optical, and mid-IR imaging and followup R $\sim$ 1000 spectroscopy of thousands of sources.
Such a census will revolutionize our understanding of the dependence of dust properties on local environment providing both an empirical description of the effects of dust on observations as well as strong constraints on dust grain and evolution models.

\pagebreak

\textbf{Background}


Dust in the interstellar medium plays a central role in star formation and galaxy evolution.
It helps shape the detailed structure of the interstellar medium (ISM), thereby directly influencing the process of star formation.
It provides crucial shielding in molecular clouds and is the main formation site for molecular hydrogen and a variety of interstellar ices, both which contribute to the rich chemistry that occurs within the dust grain ice mantles upon energetic processing as the star formation process begins.
Furthermore, the exchange of molecules between the solid and gas phase of the dense molecular cloud allows more complex species to develop in the gas than would occur in the absence of dust (i.e., the formation of more complex species such as methanol on the grains as opposed to simple molecules such as CO). 
A thorough physical understanding of interstellar dust in galaxies in the local universe is needed to better understand the properties of dust itself as well as enable a clearer picture of galaxies in general.
In addition, empirically understanding how dust extinction varies with environment enables correcting or forward modeling observations of dust extinguished objects. 

\begin{figure}[bth]
  \begin{center}
    \includegraphics[width=\textwidth]{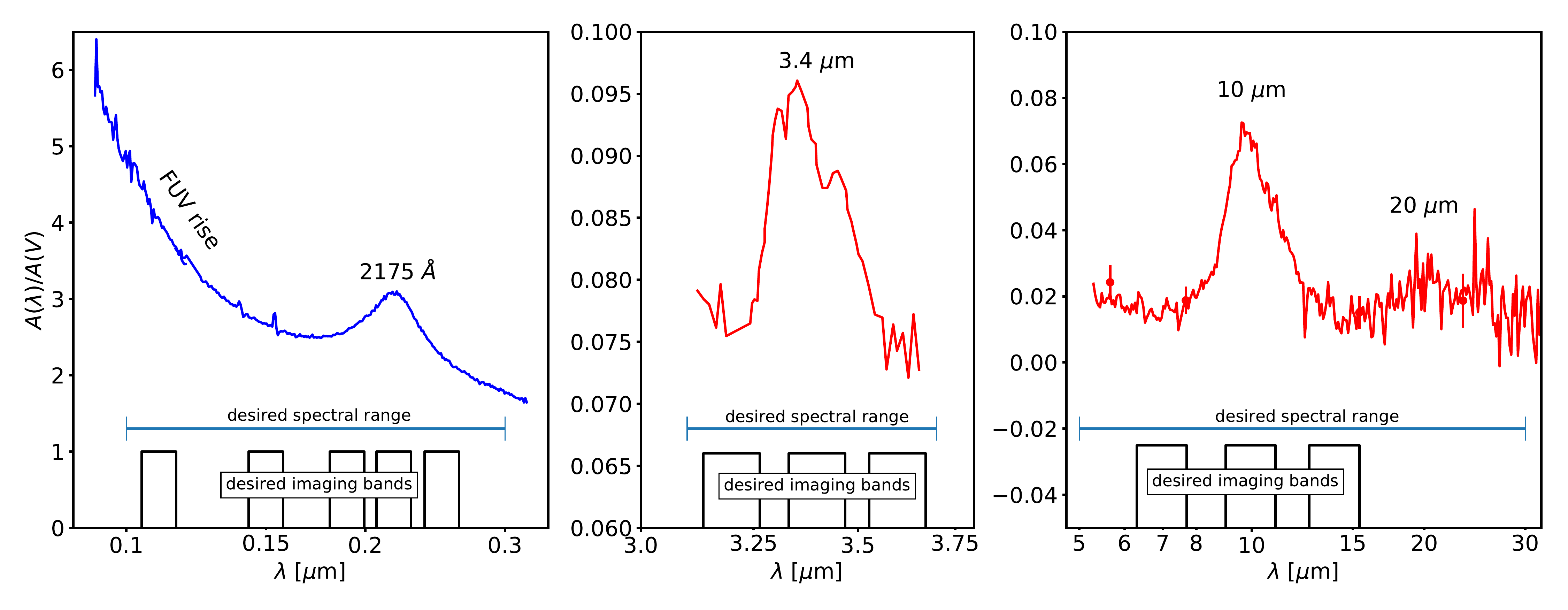}
  \end{center}
  \caption{{\small The diagnostic extinction features for carbonaceous grains (2175~\AA\ \& 3.4~\micron) and silicate grains (10 \& 18~\micron) are shown.
  The UV observations are an average of 75 Milky Way sightlines with IUE and FUSE measurements \citep{Gordon09FUSE}.
  The MIR 3.4~\micron\ measurements are for two Milky Way sightlines \citep{Pendleton02}.
  The MIR 5 -- 30~\micron\ observations are an average of $\sim$15 Milky Way sightlines with Spitzer measurements (Gordon et al.\ 2019a, in prep.).
  The needed imaging and spectroscopic observations are indicated at the bottom of the plots (see Table~\ref{tab_obs_cap}).
    \label{fig_uv_mir_extinction}}}
\end{figure}

The presence of dust is inferred from the selective scattering and absorption of light in \kempviolet{UV} and \kempred{MIR} observations; the constituents of dust have strong transitions in these wavelength regimes resulting in extinction curves (see Fig.~\ref{fig_uv_mir_extinction}) with strong features that are diagnostic of the environment and dust properties. 
These curves combine the effects of dust absorption and scattering out of the line-of-sight into a single measurement and show, among other features, the 2175~\AA\ extinction bump that has been attributed to carbonaceous dust grains \citep[e.g.,][]{Stecher65graphite, Draine84}.
In the MIR, the main extinction features are the 3.4~\micron\ feature \citep{Sandford91, Pendleton94} identified as hydrogenated amorphous carbon grains \citep[e.g,][]{Duley83, Pendleton02} and the 10 \& 18~\micron\ features identified with silicate grains \citep{Roche84, Kemper04}.
The requirement that interstellar dust be composed of highly refractory and abundant materials, combined with the plausible feature identifications described above, has lead to all successful models of interstellar dust having carbonaceous and silicate grains as their main constituents \citep[e.g.,][]{Mathis77, Weingartner01, Clayton03, Zubko04, Jones17}. 
In the Milky Way, the existing observations of these features have shown that the dust grain properties vary systematically between diffuse and dense environments providing important constraints on dust grain evolution \citep[e.g.,][]{Jones17}.
\linebreak

\begin{wrapfigure}{r}{0.5\textwidth}
  \begin{center}
    \includegraphics[width=0.5\textwidth]{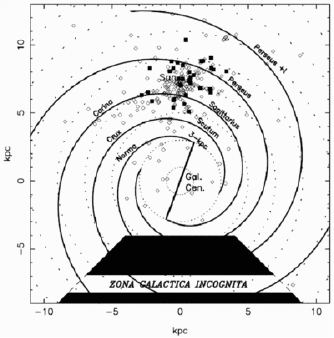}
  \end{center}
  \caption{{\small The spatial distribution of the observed UV extinction curves in the MW \citep{Valencic04}.
    \label{fig_uv_spatial}}}
\end{wrapfigure}

Our current view of dust is based, to a considerable extent, on measurements of dust extinction in the Milky Way.
\kempviolet{Currently, there exist around 400 UV extinction curves measured at spectroscopic resolution providing good constraints on the details of the 2175~\AA\ feature \citep{Valencic04, Fitzpatrick07}.}
Almost all of these MW curves can be roughly described by a single parameter $R(V)$ [$= A(V)/E(B-V)$] dependent relationship \citep{Cardelli89, Valencic04}, with a few outliers \citep{Clayton00, Valencic03}.
The spatial distribution of these MW  extinction curves is shown in Fig.~\ref{fig_uv_spatial} and clearly illustrates that our knowledge of UV dust extinction curves is limited to just the $\sim$2 kpc around the Sun's location in the Milky Way. \linebreak

\kempred{In the MIR, the number of measurements of the 10 \& 18~\micron\ silicate features along interstellar sightlines is much smaller, on the order of 20 (Gordon et al.\ 2019a in prep).
For the 3.4~\micron\ feature the number of measurements for interstellar sightlines is still smaller, on the order of just a few \citep{Pendleton02}.}
Given the small number of MIR extinction curves, the bias in our understanding of these features towards the local  Milky Way is even more extreme than for the 2175~\AA\ feature. \linebreak

\begin{wrapfigure}{r}{0.5\textwidth}
  \begin{center}
    \includegraphics[width=0.5\textwidth]{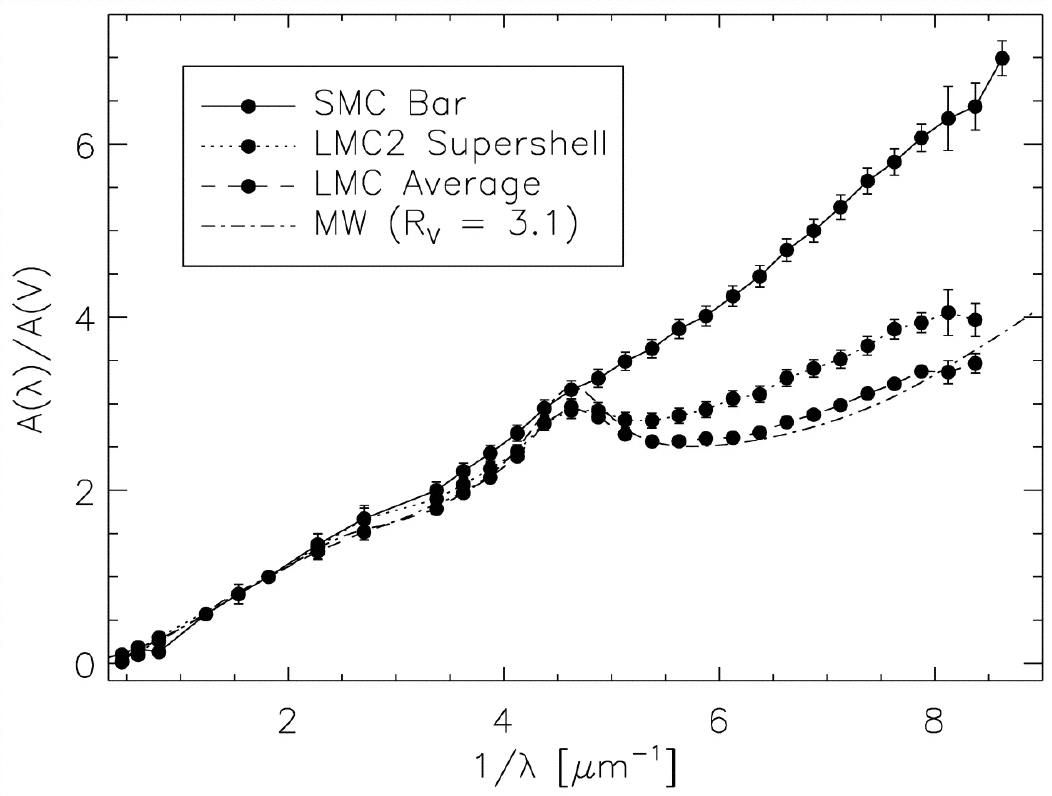}
  \end{center}
  \caption{{\small The average MW, LMC and SMC extinction curves
      are shown.  Figure from \citet{Gordon03}.
    \label{fig_all_curves}}}
\end{wrapfigure}

The Magellanic Clouds provide the nearest galaxies in which we can easily measure dust extinction at different positions through a galaxy.
\kempviolet{Due to the relative faintness of stars in these galaxies, the number of UV curves measured in both galaxies is much smaller than in the MW with 20 for the LMC \citep{Misselt99} and approximately the same for the SMC \citep[][Gordon et al. 2019b, in prep]{Gordon98, MaizAppellaniz12}}
The majority of these UV curves deviate strongly from those seen in the MW.
The most extreme variations are found in the SMC Bar where the curves have no 2175~\AA\ bump and a very steep UV slope, yet in this same galaxy there are sightlines with strong 2175~\AA\ bumps.
Fig.~\ref{fig_all_curves} shows the average LMC and SMC curves with the average curve for the
MW. \linebreak

Beyond the Magellanic Clouds, only initial probes of the UV extinction in M31 have been published \citep{Bianchi96, Clayton15}.
Ongoing HST programs in M31 and M33 (PI: Clayton) should yield samples of approximately 10 extinction curves in each galaxy. \linebreak

\kempred{There are no published extinction measurements of the MIR features along interstellar sightlines outside of the Milky Way}.
There are measurements of MIR features seen in absorption in spectra of luminous infrared galaxies \citep{Pendleton96, Wright96, Dartois04} and lines-of-sight through galaxies probed by background quasars \citep{Kulkarni11, Aller14}.
These measurements clearly show that the 3.4, 10, and 18~\micron features are present in galaxies other than the Milky Way.
But these measurements can be affected by radiative transfer effects (spectra of galaxies) or uncertainties due to the challenges of predicting the intrinsic spectrum of the background source (quasars). 
Measurements towards individual stars do not suffer from these uncertainties and can be directly connected to other line-of-sight properties (e.g., continuum extinction).
\linebreak

These extragalactic UV extinction curves illustrates that the true range of properties of dust in the Universe is larger than our current understanding of MW dust.
These observations across the faces of nearby galaxies are our current best measure that the true range of dust properties is larger than that seen in the MW neighborhood. \linebreak

\textbf{Needed Observations}

%
%

\begin{table}[tbp]
\centering
\caption{Observatonal Capabilities}
\begin{tabular}{lc} \hline
{\bf Capability} & {\bf Value} \\ \hline\hline
spatial resolution & FWHM $\sim$ 0.1'' \\
sensitivity & S/N = 100, B5V star \\ \hline
\multicolumn{2}{c}{Photometry} \\ \hline
UV central $\lambda$ & 0.11, 0.15, 0.19, 0.22, 0.25 \micron\ \\
MIR central $\lambda$ & 3.2, 3.4, 3.6, 7.0, 10.0, 14.0 \micron\ \\
areal coverage & degrees \\ \hline
\multicolumn{2}{c}{Spectroscopy} \\ \hline
spectral resolution & 500--1000 \\
UV coverage & 0.1--0.3 \micron\ \\
MIR coverage & 2.8--4.2 \& 8--30 \micron\ \\
\# spectra & 1000/galaxy \\ \hline
\end{tabular}
\label{tab_obs_cap}
\end{table}

The overall goal is to take a census of the dust extinction properties of all galaxies in the Local Group with significant dust (e.g, M31, M33, LMC, SMC, \& NGC~6822) including the Milky Way.
For the Milky Way, the main need is for MIR observations of sightlines with existing UV extinction curves and detailed studies in the UV and MIR for targeted regions probing the diffuse to dense transition.
For other Local Group galaxies, targets are needed for both the UV and MIR observations.
This can best be done using imaging observations in the UV, optical, NIR, and MIR to identify the best targets for spectroscopic observations and coarsely probe dust extinction and stellar properties.
The needed observational capabilities are summarized in Table~\ref{tab_obs_cap}, illustrated in Fig.~\ref{fig_uv_mir_extinction}, and discussed below.
\linebreak

\kemp{Sample}: Dusty Local Group galaxies span a range of metallicities from somewhat above solar (M31) to around 1/15 solar (NGC 6822) providing a wide range of environments.
The sensitivities are determined by mid-B main sequence stars as such stars probe less crowded environments than massive O stars and are more numerous.
\linebreak

\kemp{Imaging}:  
Photometric bands that probe the dust extinction features in the UV and MIR provide coarse measurements of these features for very large sample of sightlines.
High quality candidate stars for spectroscopic observations can be identified from these photometric measurements.
In addition, the feature maps across the face of galaxies can be constructed enabling a coarse characterization of dust composition within galaxies.
The UV photometric bands would include 3 bands to measure the 2175~\AA\ feature ($\lambda_c = 0.19, 0.22, 0.25$~\micron) and 2 bands in the FUV to constrain the FUV rise ($\lambda_c = 0.12, 0.15$~\micron).
The MIR photometric bands would include 6 bands, 3 to measure the 3.4~\micron\ feature ($\lambda_c = 3.0, 3.4, 3.8$~\micron) and 3 to measure the $\sim$10~\micron\ feature ($\lambda_c = 8, 10, 12$~\micron).
Measurements in optical and NIR bands is extremely useful for photometrically constraining the stellar (e.g., $T_eff$, $\log(g)$) and overall dust properties ($A(V)$, $R(V)$) as has been shown numerous times including recently in \citet{Gordon16}.
The spatial resolution needed is determined by crowding issues and is on the order of 0.1''.
\linebreak

\kemp{Spectroscopy}: The required spectroscopic capabilities would enable spectra for large samples of targets at $R \approx 500 - 1000$ resolution in the UV (0.1--0.3~\micron) and the MIR (0.3--20~\micron) to be obtained.
The spectroscopic observations would greatly benefit from multi-object capabilities over arcmin to degree field-of-views given the density of targets and galaxy sizes.
The best targets for spectroscopic observations are OB stars as they enable combined UV and MIR measurements and are intrinsically luminous.
Ideally, a large statistical sample of stars would be measured spectroscopically in each galaxy to allow for the extinction properties to be probed with environment across the face of each galaxy.
In the Milky Way, more detailed spatial studies could be done studying the variation from the diffuse to dense ISM in regions that host molecular clouds.
The UV spectroscopy would need to have high enough signal-to-noise to provide the ability to obtain UV spectral types \citep{SmithNeubig97, SmithNeubig99}.
Using ground-based telescopes to obtain classical blue-optical spectra is difficult given typical crowding of sources in many Local Group galaxies (especially M31, M33, \& NGC~6822).
Samples of 1000 sources per galaxy would be required based on our existing measurements in the MW and the desire to probe how dust properties change with environment at different galactic radii (for example).
Specifically, the sample size would provide a good sampling of the full range of dust extinction curves (50 sightlines) in broad spatial bins in each galaxy (20 spatial bins per galaxy).
For galaxies without enough OB stars, stars with later spectral types can be used to probe the MIR dust extinction features at the expense of not probing the 2175~\AA\ feature.
\linebreak

\kemp{Existing/Planned Capabilities}: 
HST provides some of the UV capabilities needed.
For imaging, the WFC3 has bands that are similar to 2 of the needed bands (F218W and F275W) and STIS has small FOV imaging in bands similar to that needed for the other two bands.
Unfortunately, the HST UV imaging capability suffers from low sensitivity and relatively small FOVs.
Regardless, these capabilities have enabled path-finding imaging in the UV for some Local Group galaxies \citep[e.g.][]{Dalcanton12, Sabbi13, Yanchulova17}.
For spectroscopy, the STIS low-resolution capabilities enable high-quality data for the entire UV spectral range at $R \sim 1000$.
These observations are time-consuming as only a single star can be observed at a time.
Path-finding UV extinction curve studies have or are being carried out using these capabilities resulting is small (5--20) sightlines in each galaxy \citep{Gordon03, Clayton15}.
\linebreak

JWST will provide a some fraction of the MIR capabilities needed.
For the 3.4~\micron\ feature, the combined NIRCam and NIRspec capabilities will provide most of what is needed for arcmin sized regions.
The NIRCam filters F300M, F335M, and F360W/F410W provide the needed 3 filters.
The NIRSpec G395M/F290LP grating combined with the multi-object capabilities of the MSA enable taking 50-100 spectra simultaneously over arcmin FOVs.
JWST will have the needed spatial resolution at wavelengths around 3.4~\micron.
Hence, JWST will provide the needed capabilities for studying this feature.
For the 10 and 18~\micron\ features, MIRI will provide the needed imaging capabilities over arcmin regions using the F770W, F1000W, and F1280W/F1500W bands.
MIRI spectroscopy for single objects will provide path-finding capabilities for the 8--20~\micron\ region, but at significantly higher spectral resolution than needed ($R \sim 2000-3000)$.
\linebreak


\kemp{New Capabilities}:

Imaging across fields many arcmin across in the UV and MIR in medium bands surrounding each feature (see Fig.~\ref{fig_uv_mir_extinction}) is critical to finding the best targets for spectroscopy and enabling truly large scale studies of the features (but limited to just coarse feature strengths).
Existing and planned facilities (inc.\ JWST) can provide large fields, but only through extensive mosaicking (e.g., PHAT).
\linebreak

Efficient spectroscopy of 1000 sources per galaxy in the UV (0.1--0.3~\micron) and MIR (between 5--30~\micron) at spectral resolutions of 500--1000 is needed to study the detailed nature of the features as their spectral shapes provide critical information on the compositions of the dust grain materials.
Having such efficient spectroscopy in the 3.4~\micron\ region would be a bonus to supplement the JWST results.
This spectroscopy could be done with multi-object capabilities (i.e., similar to NIRSpec MSA), with a large format Integral Field Unit, or with fast spectral scanning (i.e., Spitzer IRS spectral mapping).

\pagebreak

\end{document}